# Efficient Selection of Type Annotations for Performance Improvement in Gradual Typing

Senxi Li[a] , Feng Dai[a] , Tetsuro Yamazaki[a] , and Shigeru Chiba[a]

a    The University of Tokyo, Japan

**Abstract**    Gradual typing has gained popularity as a design choice for integrating static and dynamic typing within a single language. Several practical languages have adopted gradual typing to offer programmers the flexibility to annotate their programs as needed. Meanwhile there is a key challenge of unexpected performance degradation in partially typed programs. The execution speed may significantly decrease when simply adding more type annotations. Prior studies have investigated strategies of selectively adding type annotations for better performance. However, they are restricted in substantial compilation time, which impedes the practical usage.

This paper presents a new technique to select a subset of type annotations derived by type inference for improving the execution performance of gradually typed programs. The advantage of the proposal is shorter compilation time by employing a lightweight, amortized approach. It selects type annotations along the data flows, which is expected to avoid expensive runtime casts caused by a value repeatedly crossing the boundaries between untyped and typed code.

We demonstrate the applicability of our proposal, and conduct experiments to validate its effectiveness of improving the execution time on Reticulated Python. Our implementation supports a Python subset to select type annotations derived by an implemented, external type inference engine. Experiment results show that our proposal outperforms a naive strategy of using all type annotations derived by type inference among the benchmark programs. In comparison with an existing approach, the proposal achieves comparable execution speed and shows advantage of maintaining a more stable compilation time of deriving and selecting type annotations. Our results empirically indicate that the proposed technique is practical within Reticulated Python for mitigating the performance bottleneck of gradually typed programs.

ACM CCS 2012

- **Software and its engineering** → **Software design engineering**; **Runtime environments**;
- **Theory of computation** → *Type theory*;

Keywords    Gradual Typing, Performance Optimization, Static Analysis, Type Inference

## The Art, Science, and Engineering of Programming

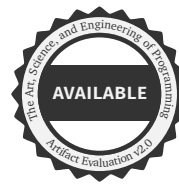 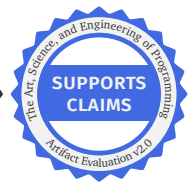

 **Introduction**

Gradual typing offers a flexible bridge between dynamically and statically typed languages. It also introduces a critical challenge of execution degradation. In a gradually typed language, programmers can add type annotations in their programs to use static typing as needed. To ensure the soundness such that a variable annotated with a type annotation $T$ must only be assigned values of type $T$, runtime casts must be performed between the boundaries between dynamically (untyped) and statically typed code. The execution degradation is often caused by these runtime casts, which check and raise an error unless values have expected types. Runtime casts are enforced in sound gradually type languages with non-erasure semantics [10, 13]. Sound gradual typing has appealed to several practical languages including Racket [22, 42], Python [43] and ActionScript [37].

A commonly used approach to improving the execution performance is to automatically add additional type annotations by type inference so that more variables will be statically typed. However, naively adding all type annotations suggested by type inference without careful consideration may unexpectedly cause execution degradation. This is because adding some type annotations may cause values repeatedly crossing the boundaries between untyped and typed code, which would introduce extra runtime casts. For example, when an untyped value is assigned to a variable $v$ and $v$ is an argument to an untyped function parameter, adding an additional type annotation to $v$ would let a value cross the boundaries from untyped to typed, and then back to untyped. Another commonly used approach is to leverage Just-In-Time (JIT) compilation to avoid redundant runtime casts at runtime. While JIT has been an effective approach for performance improvement in gradual typing [3, 38, 39], its applicability is limited for systems with sufficient resources since JIT requires additional memory by dynamic compilation. For example, microcontrollers such as MicroPython [11] and other embedded systems often lack JIT support due to their constrained memory.

Research studies have revealed that it does not reliably lead to performance improvement by naively adding more type annotations. In particular, Takikawa et al. report non-negligible performance degradation when a program is partially typed [41]. Their work shows that unless a program is entirely statically typed, there would be significant execution slowdown compared to the untyped program such as more than 100 times even if the program is mostly typed. The performance pitfall discourages the usage of gradual typing since an easy workaround is to just remove all type annotations.

A few existing researches have explored approaches to select added type annotations from the ones derived by type inference for improving the performance of gradually typed programs. However, they may incur non-negligible compilation time for deriving and selecting type annotations. For instance, we observe that one existing tool Herder [6] would incur a significant amount of compilation time such as more than 10 minutes for several benchmark programs, though they are relatively small in lines of code compared to real-world programs. It remains challenging to select effective type annotations for improving the performance considering practical usage.

We propose a lightweight method, named *TypePycker*, to selectively add type annotations for improving the execution time of gradually typed programs. The novelty of TypePycker compared to existing approaches is its lightweight method that achieves comparable execution performance but does not require long compilation time. Given a program, it performs type inference and selects only a subset of the type annotations derived by the type inference. To avoid significant costs of runtime casts caused by values repeatedly crossing the boundaries between untyped and typed code, TypePycker uses a quick, amortized approach to select type annotations placed along the data flows.

To evaluate the effectiveness of the proposed method, we conduct experiments on a set of collected programs in a gradual typing language Reticulated Python [43]. Our experiment results show that the proposed method achieves performance improvement compared to naively using all inferred types derived by type inference. Across 41 collected benchmark programs, our method outperforms using all inferred type annotations in 32 programs. It achieves a speedup up to more than 5x compared to a program appended with all inferred type annotations. Our method gives comparable execution time against using all inferred type annotations in 6 programs. For assessing the compilation time, we compare TypePycker with the existing tool Herder [6]. We observe that TypePycker maintains more stable and acceptable compilation time, and also achieve comparable execution time across the benchmark programs. We also find that TypePycker offers clear advantages in several programs such that it maintains much faster compilation time than Herder while at the same time our selection gives better performance than using all inferred types. Our findings show that the proposal is a practical method to mitigate the performance degradation of gradually typed programs in Reticulated Python.

The rest of the paper is organized as follows. Section 2 motivates the reader by introducing the performance challenges in gradual typing. Section 3 presents the proposed method. Section 4 describes our experiments. Section 5 discusses threats of validity. Section 6 reviews related works and a brief conclusion ends this paper.

## 2 Performance Degradation in Gradually Typed Programs

Predicting the execution performance of gradually typed programs is challenging. The reason is that a gradually typed language would perform runtime casts when a value crosses the boundary between untyped and typed code. This often causes serious degradation of execution performance. A runtime cast is an operation executed at runtime to check if a value has an expected type. Runtime casts are executed for ensuring type invariants specified by the program [7, 37, 40]. A number of gradually typed languages perform runtime casts for sound static types [22, 33, 43]. Several researches have observed significant runtime overhead caused by runtime casts where a partially typed program is more than 100 times slower than its untyped version [15, 41].

Consider an example program in a gradually typed language given in Listing 1. A local variable z is annotated with a type annotation by the programmer. A local variable, function parameter or return type that without a type annotation is implicitly typed as the unknown type *. We assume that succ is a library function typed as Function([*], *), which denotes a function type with a parameter type * and return type *.

◼ **Listing 1** An example program. succ is assumed to be typed as Function([*], *). * denotes the unknown type.

```
1  def f(x, y):
2      u = x
3      z: Bool = y
4      return if z then succ(x) else succ(u)
5
6  f(succ(1), true)
```

The runtime casts for the example program are presented in Listing 2. Note that the language implicitly gives static types to literals such as 1 and true. This makes the program partially typed and involve runtime casts even when the program does not include any explicit type annotations. For illustration purposes, we use the notation $\{T_2 \Leftarrow T_1\}e$ to explicitly show the occurrence of a runtime cast. It reads as expression $e$ has type $T_1$ at compile time while $e$ is expected to have type $T_2$ at runtime. For example, a runtime cast $\{\text{Bool} \Leftarrow *\}y$ occurs at line 12 since y is implicitly typed as * but it is assigned to a local variable typed as Bool. This runtime cast is performed at runtime to check whether y is a boolean value, which causes runtime penalty.

◼ **Listing 2** Example program with runtime casts. succ is assumed to be typed as Function([*], *).

```
10  def f(x, y):
11      u = x
12      z: Bool = {Bool ⇐ *}y
13      return if z then succ(x) else succ(u)
14
15  f(succ({* ⇐ Int}1), {* ⇐ Bool}true)
```

A widely used approach to mitigating the execution slowdown due to runtime casts in gradual typing is to automatically add additional type annotations derived by type inference so that a number of runtime casts will be eliminated. Suppose that a type inference engine is run to derive additional type annotations for local variables, function parameter, and return types for the example program in Listing 1. It infers the type of parameter y as type Bool. This additional type annotation is expected to improve the execution speed since it can eliminate a runtime cast $\{* \Leftarrow \text{Bool}\}y$ at line 15 and a runtime cast $\{\text{Bool} \Leftarrow *\}y$ at line 12 shown in Listing 2.

However, adding type annotations as recommended by the type inference engine, without careful consideration, may lead to performance degradation that contradicts programmers' intuitive expectations. Suppose that an additional type annotation Int is added to parameter x in Listing 1. The runtime casts would occur as

```
1  def f(x: Int, y):
2      u = {* ⇐ Int}x
3      z: Bool = {Bool ⇐ *}y
4      return if z then succ({* ⇐ Int}x) else succ(u)
5
6  f({Int ⇐ *}succ({* ⇐ Int}1), {* ⇐ Bool}true)
```

Compared to the runtime casts presented in Listing 2, three extra runtime casts occur since a value repeatedly crosses the boundaries from untyped code succ(1), to typed code x, then back to untyped code u and the argument in succ(x). Recall that succ is assumed to be typed as Function([*], *).

The unexpected performance degradation caused by excessive type annotations is well-known by existing studies. Takikawa et al.'s seminal paper [41], for instance, shows that unless a program is entirely statically typed, the execution time may be slow even if the program is mostly typed. We also observe that simply adding more additional type annotations derived by type inference does not necessarily lead to better performance. We does similar experiments to those by Takikawa et al. as our preliminary study. In our preliminary study, we measure the execution time of several public Python programs annotated with different numbers of type annotations on Reticulated Python [43]. Reticulated Python is a gradually typed dialect of Python that allows programmers to annotate programs with static types. It performs runtime casts for sound static types. Given a Python program, we give a set of type annotations for its local variables, function parameter, and return types. Those type annotations are derived by an external type inference engine. Note that we modify Reticulated Python to support static types for local variables. The original Reticulated Python does not support explicit type annotations for local variables. Then we generate variants of the program annotated with random subsets of the type annotations. We measure the execution time of the program with zero number of type annotations, the program including all the type annotations derived by type inference, and the generated program variants with different subsets of those type annotations.

Figure 1 demonstrates the results for two of the tested programs.[1] For the program in Figure 1a, the program including all the type annotations derived by type inference gives better performance than the program with zero number of type annotations. This showcases that adding type annotations can improve the performance. However, it is also observed that the performance can be further improved with fewer type annotations. For the program in Figure 1b, although type inference generates type annotations for more than 80% of variables and others, the program that includes all of these annotations performs worse than the program without any annotations. The execution time can be improved by discarding some of the type annotations.

An interesting research direction is to develop techniques for selecting appropriate type annotations, from those derived by type inference, that can improve execution performance. However, existing approaches have limitations. One study [6] suggests appropriate type annotations by analyzing the overhead of runtime casts with different subsets of possible type annotations. A drawback of this approach is that it consumes extremely long compilation time for several benchmark programs including real-world ones. In this paper, the compilation time means the time for deriving additional type annotations by type inference and inserting selected type annotations into the original program. Another recent study [21] tries to tackle this problem by predicting the optimal execution performance with a machine-learning based method. However,

---

[1] All the other results are included in our artifact.

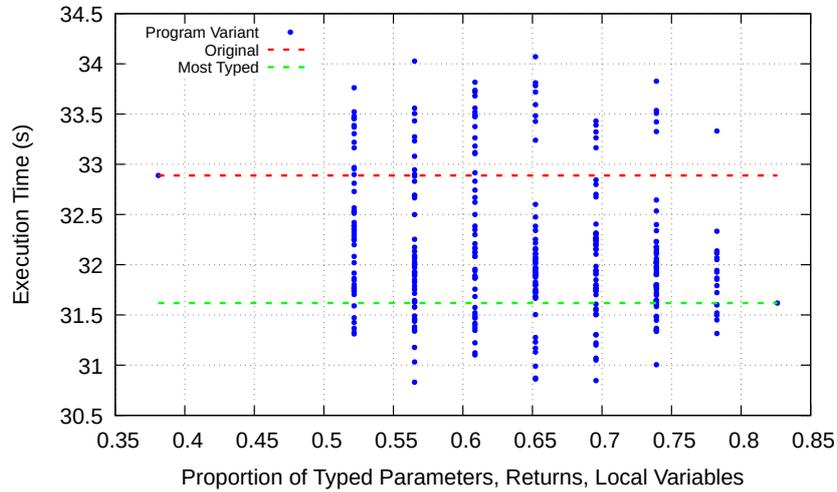

**(a)** storage

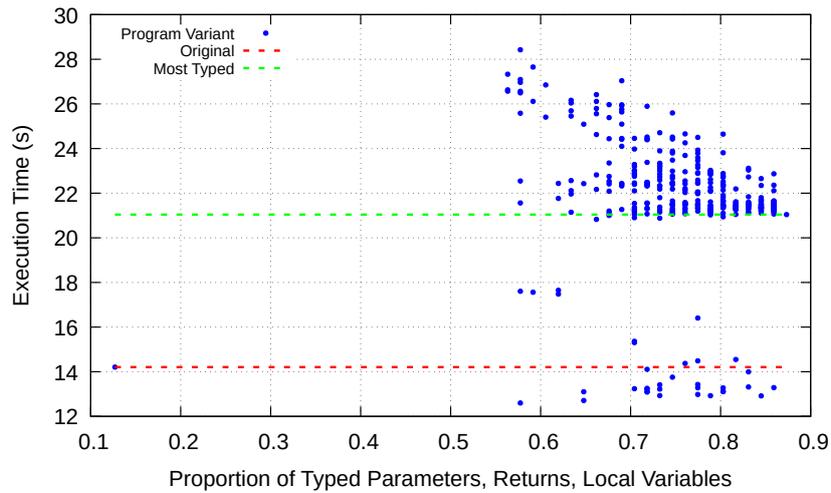

**(b)** nbody

■ **Figure 1**  Execution time for Python program variants. Each blue dot is a generated program variant with a different subset of type annotations. X-axis shows the proportion of typed function parameters, returns, and local variables. Y-axis shows execution time in seconds (lower is better). The red dashed line represents the program with zero number of type annotations. The green dashed line represents the program including all the type annotations derived by type inference.

it requires the execution of a number of programs with different subsets of type annotations for training. As a result, it remains challenging to select candidates of type annotations considering practical usage.

## 3 Efficiently Selecting Added Type Annotations

We propose TypePycker, a new lightweight method for improving the execution time of gradually typed programs by selectively adding type annotations. Our method first performs type inference and selects only a subset of the type annotations derived by the type inference so that the program including the selected type annotations will run faster than the program before the type annotations are added. Unlike previous methods, TypePycker is lightweight and hence does not require long selection time. Despite being lightweight, our method shows comparable performance improvements, and in some cases, even outperforms an existing tool, according to our experiments.

Our key idea is to select type annotations based on their positions on the data flows. The costs of runtime casts tend to increase when a value is passed from a typed variable to an untyped (or unknown-type) variable, and then back to a typed variable, repeatedly. Thus, TypePycker selects the inferred type annotation for a variable $v$ when type inference infers the types for all the other untyped variables (and function calls) that can reach $v$ along the data flows. Our method quickly checks this condition using a lightweight, amortized approach. It does not repeatedly perform expensive traversal through data flows, nor enumerate all possible occurrences of runtime casts in a brute-force manner.

### 3.1 SimpliPy

To present the proposed method, we begin by defining a small gradually typed language called *SimpliPy*. SimpliPy serves as a simplified, illustrative version of Python. We will use SimpliPy to present the proposal for simplicity.

■ **Listing 3** Language syntax

```
1  prog := stmt
2  stmt := expr | returnStmt | defStmt
3        | var = expr | expr[expr] = expr
4        | stmt; stmt
5  returnStmt := return expr
6  defStmt = def id(param, ...) -> T: stmt
7  var, param := id: T
8  expr := expr + expr | expr(expr, ...)
9        | if expr then expr else expr
10       | [expr, ...] | expr[expr]
11       | id | number | true | false
12
13 T := * | Bool | Int | Array(T) | Function([T, ...], T)
```

The syntax of SimpliPy is given in Listing 3. A program is a sequence of statements, and an expression supports addition, arrays, conditional, and function application.

A type is either boolean, integer, function, array, or unknown. The unknown type is denoted by *. An expression of the unknown type is often called *an untyped expression*. [expr, ...] and (expr, ...) represents a sequence consisting of zero or more instances of expr, separated by commas and surrounded with brackets [] and (), respectively. [expr, ...] is an array literal. In SimpliPy, programmers may not omit a type annotation. Instead, they explicitly specify the unknown type using the type annotation : *. When a concrete type annotation : Int is substituted for : *, however, we below refer to this as adding or appending a type annotation to an untyped variable or parameter for the sake of readability.

## 3.2 Graph Construction

The proposed method first constructs a directed graph for a given SimpliPy program. This graph represents data flows potentially affecting runtime casts for gradual typing. Our graph is statically constructed based on the syntax of a given program. It also assumes that a type-inference engine generates inferred types for the given program.

A vertex represents a variable, a function parameter, a function name, a literal, or an expression. When an expression contains sub-expressions, all the expressions are represented by vertices. For example, an expression x + y generates three vertices for x, y, and x + y. An expression a[i] generates three vertices for a, i, and a[i].

Each distinct variable, parameter, or function name, corresponds to one single vertex, regardless of the number of occurrences within the program. Their different occurrences are represented by one single vertex if they are bound to the same variable, parameter or function name in the program's name scope determined by static name resolution. Furthermore, every function definition generates a vertex for representing a return statement. A function definition generates a single vertex even if it contains multiple return statements.

Each vertex has two properties. One property is a given type. We represent the given type for a vertex $v$ as $given(v)$. If a vertex represents a variable or a function parameter, its given type is the type given by a type annotation written by programmers. For example, for x: Int = 3, the given type for the vertex representing x is Int. For x: * = 3, the given type for the vertex representing x is *. If a vertex represents a return statement, its given type is the return type given by the type annotation for its function definition. If a vertex represents a function name, its given type is the function type given by its type annotations. If a vertex represents a number or boolean literal, then its given type is Int and Bool, respectively. If a vertex represents an empty array literal [], its given type is Array(*). If a vertex represents a non-empty array literal such as [x, y], its given type is Array(*). The given types of the other vertices are *. Note that some vertices, such as ones for number literals, are not associated with type annotations written by programmers, yet their given types are not *.

The other property of a vertex is an inferred type. We represent the inferred type for a vertex $v$ as $infer(v)$. We assume that an external type inference engine is run and the resulting inferred types are reflected on the graph. For example, when the type of an expression f(x) is inferred as type Int, then Int is the inferred type for the vertex representing that function application. The type inference engine may infer the types of other kinds of expressions. These type are used as the inferred types of the vertices corresponding to those expressions. If the type inference engine does not infer any type, then unknown type * is considered as the inferred type. The inferred type for the vertex representing a `return` statement is the type inferred as the return type of its function definition. The inferred type for the vertex representing a function name is the function type inferred for that function. The inferred types for number and boolean literals are their value types, `Int` and `Bool`. We assume that *infer(v)* is more concrete than *given(v)*, meaning that at least one occurrence of * in *given(v)*, if existing, is replaced with a concrete type in *infer(v)*, as a type-inference engine will consider *given(v)* when computing *infer(v)*. For example, if a type inference engine could not derive additional type information for *v*, then *infer(v)* would be equivalent to *given(v)*.

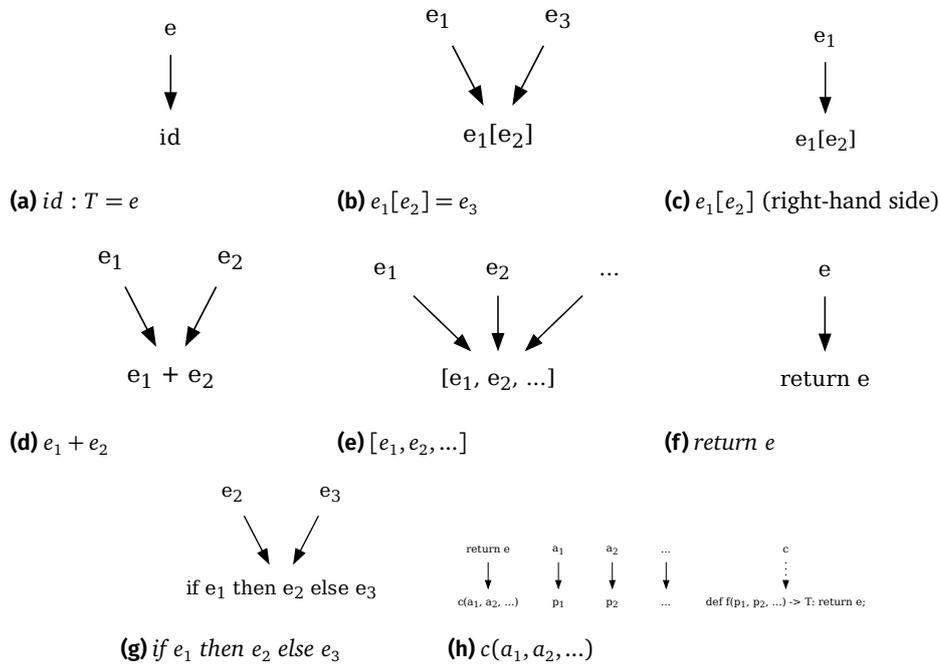

**(a)** *id : T = e*     **(b)** $e_1[e_2] = e_3$     **(c)** $e_1[e_2]$ (right-hand side)

**(d)** $e_1 + e_2$     **(e)** $[e_1, e_2, ...]$     **(f)** *return e*

**(g)** *if $e_1$ then $e_2$ else $e_3$*     **(h)** $c(a_1, a_2, ...)$

**■ Figure 2** Edge creation. Solid arrows denote graph edges. Dashed arrows indicate callee functions discovered via a points-to analysis.

A directed edge is created to represent a data flow in a program. Figure 2 presents how directed edges are created. The edge creation largely follows standard practices in data flow analysis. For a function application $c(a_1, a_2, ...)$, edges are created based on the potential callee functions that *c* may refer to. To discover a callee function, we run an arbitrary points-to analysis. When the analysis discovers multiple potential callee functions, edges are created between the function application and every potential callee function. When the analysis does not discover any callee function, no edge is created. The point-to analysis does not need to be highly precise. Imprecise analysis would simply result in excessive type annotations, which could potentially degrade runtime performance.

### 3.3 Selecting Inferred Types

After the graph construction, the proposed method determines whether an inferred type is appended to the original source code as a type annotation.

First, our proposed method selects the vertices $v$ that represent either a variable, a parameter, or a return statement for which $given(v)$ is a type containing *. A type contains * if it is *, an array type containing elements of such a type, or a function type including such a type as a parameter type or a return type. Array(*) and Function([*],Int) are examples. These selected vertices $v$ are candidates that new type annotations are appended to, based on $infer(v)$. Recall that we assume that $infer(v)$ is always more concrete than $given(v)$.

Then, for each of the selected vertices $v$, it enumerates all the *closest source vertices* reachable to $v$. If no closest source vertex exists or all the closest source vertices $w$ hold that $given(w)$ is not *, then $infer(v)$ is appended as a type annotation for $v$. Here, a source vertex is a vertex without an incoming edge or a vertex $w$ holding that $given(w)$ does not contain *. A source vertex $w$ is a closest source vertex reachable to $v$ if there exists a path from $w$ to $v$ and that path does not include a source vertex except for $v$ and $w$ ($w \neq v$).

Consider the example program in Listing 1. First, vertices representing variable u, parameter x and y, and the return statement of function f are selected as candidates. The vertex representing variable z is not selected as a candidate since its given type is Bool, which does not contain *. For the vertex representing y, its closest source vertex is the vertex representing true. The inferred type for y is appended as a type annotation since the given type for true is not *. For the vertices representing x and u, their closest source vertex is the vertex representing succ(1). For the vertex representing the return statement of function f, its closest source vertices are vertices representing succ(x) and succ(u). The inferred types for x, u and the return type of function f are not appended as additional type annotations since the given types for the function applications of succ are *. Recall that function succ is typed as Function([*], *).

### 3.4 Correctness

TypePycker's correctness is guaranteed by a correct type inference engine and the underlying gradually typed language. The correctness is defined as: an optimized program with additional type annotations must either produce the same value or fail at the same location as the original program [37]. This correctness follows the *gradual guarantee* [40] of a gradually typed language, that is, a gradually typed program is still well-typed or correctly executed when discarding some of its type annotations. Since TypePycker only appends a type annotation that is either $infer(v)$ or $given(v)$, a selection of the additional type annotations by TypePycker is equivalent to discarding some type annotations of a gradually typed program whose type annotations are $infer(v)$. If the type inference engine only derives correct additional type annotations, then TypePycker is also guaranteed to be correct.

## 4 Experiment

We conduct experiments on a set of collected benchmark programs using a gradually typed language Reticulated Python. The goal of our experiments is to verify whether the proposed method can improve the execution time of gradually typed programs while maintaining a stable and acceptable compilation time of deriving and selecting additional type annotations.

In the following experiments, we first present that our method improves the execution time. We compare the execution time of each benchmark program appended with all additional type annotations by type inference, against that appended with only selected type annotations by our method. We then show that our method TypePycker achieves a more stable compilation time than an existing work Herder [6]. We also show that TypePycker can obtain comparable execution time compared to Herder. Furthermore, we demonstrate that our method provides performance gains across different implementations of Reticulated Python. Additionally, we test several benchmark programs on Reticulated Python with PyPy to showcase whether TypePycker is possibly beneficial to a gradually typed language with JIT compilation.

### 4.1 Implementation and Dataset

Our configuration for the experiments consists of a compiler that appends type annotations to a program, and an extension over Reticulated Python for running the compiled program. To derive additional type annotations, we implement a constraint-based type inference engine [36] for a Python subset. We utilize an existing compiler toolkit *InferType* [26]. InferType is an embedded domain-specific language in Java for implementing constraint-based type inference. Our engine supports inferring types for functions, local variables, class methods and class fields. Given a Python program, our type inference engine assigns a unique type variable to each expression, and collects type constraints over types and type variables by traversing the AST of the program. It then computes a binding that associates type variables to their inferred types by solving the collected type constraints. InferType invokes an SMT solver for computing that binding. We omit a detailed description of the adapted type system for brevity. Comprehensive details on the typing rules can be found in existing studies [16, 20].

We implement the proposed method on a Python subset. We extend the presented techniques in SimpliPy to Reticulated Python by following common data flow analysis on a graph-based representation. For example, the graph construction for other binary operations in Reticulated Python is similar to that presented in SimpliPy. To identify potential callee functions of function applications in the graph construction, we adopt a context-insensitive, flow-insensitive points-to analysis. The analysis approximates a set of callee functions for each function application without considering conditional branches or the order of execution. For example, consider the following code fragment:

```
1  e = f
2  e(true)
3  e = if cond then h else g
4  e(9)
```

where f, g and h are identifiers of three function definitions. The analysis would compute that e at both line 2 and line 4 may refer to either function f, g or h though true is never passed to h or g, and 9 is never passed to f during execution. It would create edges from arguments true and 9 to the parameters of f, g and h in the graph construction as we described in Section 3. Despite its simplicity and imprecision, the adopted points-to analysis is practical for improving the performance among the benchmark programs in our experiments.

■ **Listing 4**  Example program with generated fast function

```
1   def f_fast(x: *, y: Bool):
2       u = x
3       z: Bool = y
4       return if z then succ(x) else succ(u)
5
6   def f(x, y):
7       u = x
8       z: Bool = y
9       return if z then succ(x) else succ(u)
10
11  f_fast(succ(1), true)
```

Our implementation extends Reticulated Python with two main modifications. First, we adapt a technique from a recent study [7]. The goal of adapting this technique is to preserve program behaviors while maximizing optimization opportunities. Preserving program behaviors here means that optimized programs must either produce the same value or fail at the same location as the original program [37]. The main idea is to generate an optimized version of each function appended with additional type annotations while also keeping the original function. It then dispatches function applications based on the static types of arguments and parameters. For example, the example program in Listing 1 with additional type annotations will be generated as shown in Listing 4. An optimized function called f_fast is generated. The original function is also kept in the source code. In the main part of program, the optimized function is called because the static types of the arguments are subtype of the additional parameter types. Below we refer to this technique as *fast-slow*.

Second, we extend Reticulated Python to support type annotations for local variables. This enables us to select additional type annotations for local variables as we described in SimpliPy. The original Reticulated Python does not support explicit type annotations for local variables.

We build a dataset for the experiments. We make it publicly available in our artifact. The programs in the dataset are not annotated with explicit type annotations for function parameters, function returns or local variables by programmers. Note that although we use those un-annotated programs as baseline in the experiments, the programs are not entirely untyped. The programs are partially typed since the underlying language Reticulated Python always gives static types to several expressions such as number and array literals. This partial typing introduces non-trivial scenarios with runtime casts where there are opportunities to append or discard additional type annotations for improving the execution performance. Thus, although the programs are not ideally annotated with explicit type annotations by programmers, our dataset is qualified as a testbed for experimenting if the proposal can improve the execution performance.

The dataset contains 50 Python programs collected from several public sources. 31 programs are collected from several academic studies [6, 26, 28]. Those programs include a subset of the Python performance benchmark suite,[2] and programs that are used to evaluate Reticulated Python used by existing researches [5, 7, 21, 43]. Below we refer to these programs as *MicroBench*. 6 programs are collected from a Python textbook for educating undergraduate students used at The University of Tokyo.[3] Although they were not designed or used for gradual typing, we assume that they exhibit realistic Python coding patterns in education and serve as representative real-world programs.

9 programs originate from the book Structures and Interpretation of Computer Programs (SICP) [1]. Those programs are expected to be challenging in terms of compilation time for type inference and also statically constructing, analyzing data flow graphs, as they contain a larger number of function definitions and function applications (presented in Table 1) including the use of function pointers. Since the original programs are written in Scheme,[4] we utilized a large language model (Gemini 2.5 Flash) to help translate them to Python. We used the following prompt appended with an original Scheme program for the translation:

```
1  Translate this program to Python. This program is adapted from the book
2  Structures and Interpretation of Computer Programs. Include a main
3  function for execution. Do not use f-string, lambda expression, starred
4  expression, variable argument or class.
```

Due to the limitations in Reticulated Python and Herder, several Python features such as classes are excluded during the translation, as they are not used in the resulting programs.

The other 4 programs are modified variants with nested function calls of 4 programs from MicroBench. These programs are designed to test the compilation time and performance impact of our method for programs with nested function calls. Here, a nested function call refers to a function call within another function. Given a program, we manually refactor complex or long statements to use helper functions. We create 3 to 6 helper functions for each program. For example, a code fragment from one benchmark program is given as

```
1  def SOR_execute(omega, G, cycles):
2      w,h,d = G
3      for p in range(cycles):
4          for y in range(1, h - 1):
5              ...
```

We refactor it into

---

```
1  def SOR_execute(omega, G, cycles):
2      w,h,d = G
3      for p in range(cycles):
4          loop_heights(p, w, h, d, omega, G)
5
6  def loop_heights(p, w, h, d, omega, G):
7      for y in range(1, h - 1):
8          ...
```

loop_heights is a manually created helper function and it is called within function SOR_execute, which introduces a nested function call. Below we refer to these manually synthesized programs as *Syn*. All code changes do not change the functionality of the original programs. A complete list of all code changes is included in our artifact.

Table 1 outlines some detailed characteristics of the benchmark programs. Each row in the table categorizes the dataset by its data source. Although our dataset is not exceedingly large and may introduce some collection bias, it includes more programs from multiple public sources, compared to other datasets used in existing studies [6, 7, 21, 44].

■ **Table 1**  Dataset overview. Programs are grouped by data source (number of included programs). Each cell gives min/mean/max values in the programs. Def and Call are numbers of lexical function (and method) definitions and function applications. Edge is number of edges in the constructed graph by TypePycker.

| Group (Program) | LOC | Def | Call | Edge |
|---|---|---|---|---|
| MicroBench (31) | 22/64/145 | 2/7/15 | 2/11/25 | 51/319/823 |
| PyTextbook (6) | 55/99/215 | 7/12/25 | 13/25/48 | 259/489/1011 |
| SICP (9) | 181/317/504 | 19/42/75 | 18/56/105 | 509/1052/1760 |
| Syn (4) | 55/85/139 | 9/11/12 | 22/30/38 | 390/500/804 |

The experiments are performed on a Ubuntu 24.04.2 LTS desktop machine equipped with an AMD Ryzen 9 9950X 16-Core Processor. The desktop machine is installed with OpenJDK 17.0.5, Python 3.9.7 for our implementation. The programs are executed using the guarded semantics of Reticulated Python [43, 44].

## 4.2  Validating Selection Effectiveness

We compare the execution time of the benchmark programs to examine if the proposed selection of additional type annotations can improve the performance of gradually typed programs. Given a benchmark program, we compare the execution times of its three variants: a *Given* variant that is equivalent to the original benchmark program; an *Infer* variant that includes all additional type annotations derived by type inference; a *Chosen* variant that includes only selected additional type annotations by TypePycker. We execute each of the program variants for 10 times and measured the arithmetic mean of the elapsed time. Below we use the terms Given, Infer and Chosen to refer to these three kinds of programs.

Our method selects all the additional type annotations derived by type inference in 9 of 50 benchmark programs. This is an expected behavior of our method. Intuitively, if a gradually typed program can be enhanced by a type inference engine to be fully typed, or if all additional type annotations are expected to eliminate runtime casts, one should use those additional type annotations for optimal performance instead of discarding them. For example, one benchmark `Mandelbrot` from MicroBench calculates the classic fractal, which mainly involves binary operations over floating-point numbers. It does not use any unknown library functions with respect to types. Our type inference engine infers most of the types within the program. Below we exclude the results of these 9 programs from further comparisons in this subsection. Infer and Chosen for these programs also give close performance in the experiments.

Experiment results show that the proposed method gives positive performance impact in the benchmark programs. Figure 3 presents the execution times of Given, Infer, and Chosen for each benchmark program. To qualitatively assess the performance impact, we categorize the relative execution time of Infer and Chosen for each benchmark program into three cases. The categorization is based on the arithmetic mean (*Mean*) and standard error (*SE*) of the measured execution times:

$$\text{case} = \begin{cases} \text{win} & Mean_{Chosen} + SE_{Chosen} < Mean_{Infer} - SE_{Infer} \\ \text{loss} & Mean_{Chosen} - SE_{Chosen} > Mean_{Infer} + SE_{Infer} \\ \text{tie} & \text{otherwise} \end{cases}$$

A program is considered as either a *win* or *loss* case if the error bars of Infer and Chosen do not overlap. Otherwise, it is considered as a *tie* case. This indicates that the performance impact by our method is not significant.

Our method gives win cases in 32 of 41 programs, and gives tie cases in 6 of 41 programs. Figure 3e presents the speedup ratio comparing Infer and Chosen, which is calculated as $Mean_{Infer}/Mean_{Chosen}$. Chosen outperforms Infer with a more than 1.1x speedup ratio in 22 programs, and with a more than 5x speedup ratio in 4 programs. Infer is slower than Given, while Chosen is faster than both Given and Infer in 10 programs. This demonstrates scenarios when simply appending all the type annotations suggested by type inference degrades the performance of gradually typed programs, but the selection of them can lead to performance improvement.

It is also observed that our method does not always improve the performance. It gives loss cases in 3 programs. We manually inspect the slower programs and find that the performance degradation is mainly due to the limitation of static analysis. In dijkstra_alternate from Figure 3a, the code fragment is illustrated below:

```
1  u = expr
2  for v in vertices:
3      distances[v] = distances[u] + graph[u][v]
```

Variable `u` is assigned to `expr` that is typed as the unknown type `*`, and `u` is used as an index for array accesses later, which is expected to be type `Int`. The type for `u` is inferred as `Int`, which is appended in Infer but not appended in Chosen. This changes the occurrence of a runtime cast `Int ⇐ *` from the assignment to the later array accesses. However, the later runtime casts occur in the for loop, which causes

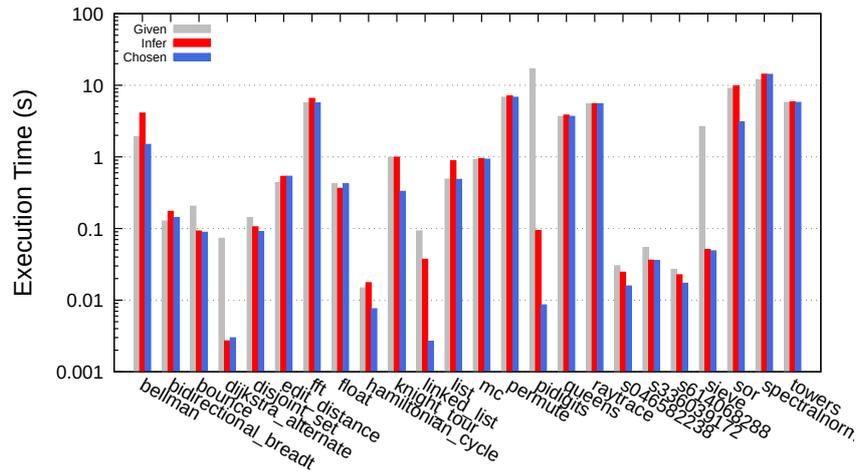

**(a)** MicroBench

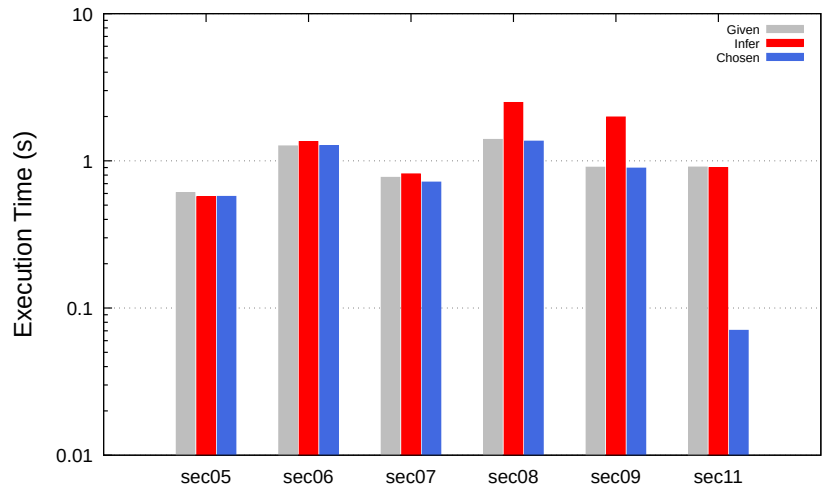

**(b)** PyTextbook

■ **Figure 3** Execution time results (Part I).

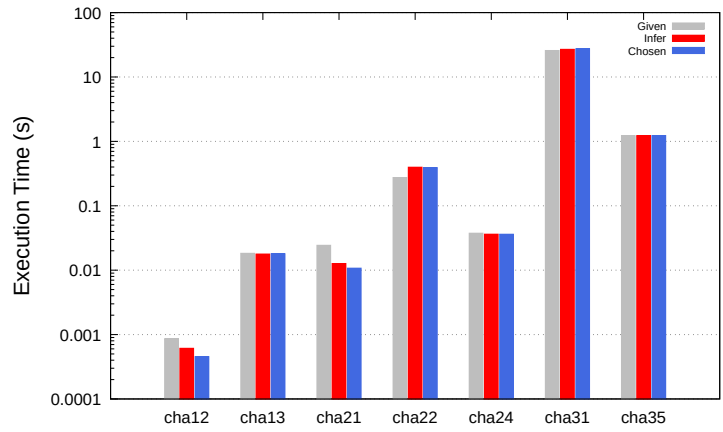

**(c)** SICP

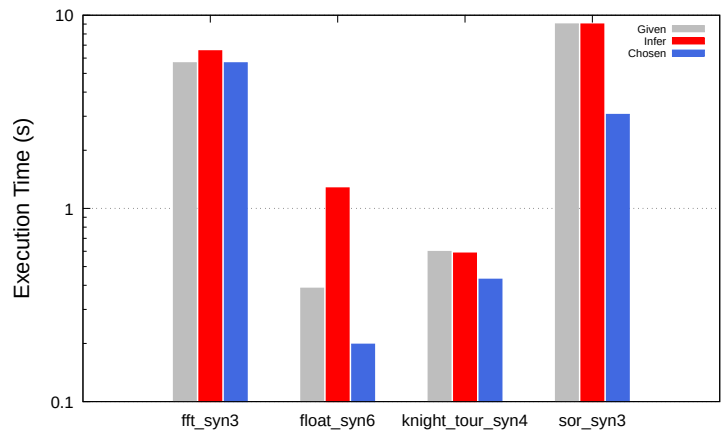

**(d)** Syn

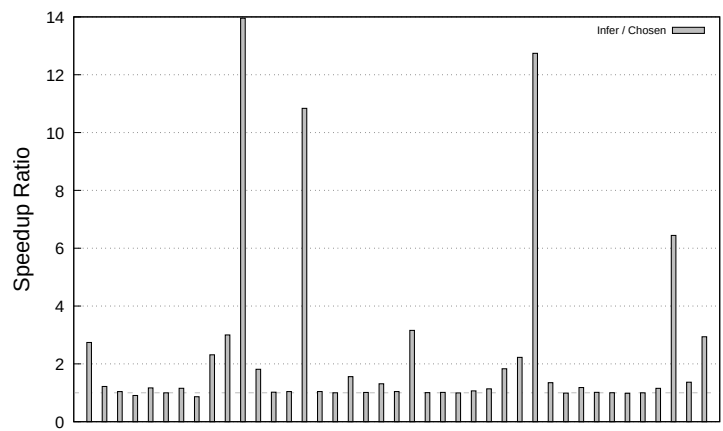

**(e)** Speedup ratio (Infer over Chosen)

■ **Figure 3** Execution time results (Part II). (a)-(d): Each sub-figure group programs by data source; X-axis shows program names; Y-axis shows execution time in seconds; Each program features Given, Infer and Chosen from left to right; Each bar is average of 10 runs; Lower is better. (e): Speedup ratio of Infer over Chosen; X-axis shows programs in (a)-(d); > 1 is better.

more overhead during execution in Chosen. Another slower case in program float has a similar scenario. For the other slower case cha31, Chosen appends different type annotations and produces multiple different occurrences of runtime casts from those by Infer. It is difficult to conclude the slowdown into a straightforward reason.

We also inspect and find that our method may not give better performance when runtime casts would occur in different control-flow branches. For example, one code fragment in program seco8 is shown below:

```
1  def mergesort(xs):
2      if len(xs) <= 1:
3          return xs
4      else:
5          return merge(...)
6
7  mergesort([randint(0, 100) for _ in range(70)])
```

The argument at line 7 is an array literal typed as Array(*) since randint is a library function whose type is not available. The return type of function merge is Array(Int) in both Infer and Chosen. The return type of mergesort is inferred as Array(Int). In Infer, this return type annotation would incur an occurrence of a runtime cast Array(*) ⇐ Array(Int) in the if branch, but no runtime cast in the else branch. In Chosen, a runtime cast * ⇐ Array(*) occurs in the else branch. TypePycker does not select the return type annotation for mergesort since one of its closest source vertices (representing the array literal at line 7) contains *. However, the program executes the else branch more frequently, which results in less runtime overhead in the else branch in Infer. The selection for this function return type degrades the execution performance.

### 4.3 Demonstrating Stable Compilation Time

We compare our method with an existing work Herder to show that our method is more stable and practical with respect to compilation time. Here, the compilation time mainly includes type inference for deriving additional type annotations and inserting selected type annotations into the original program. Herder [6] is a tool that statically analyzes the cost of runtime casts of a gradually typed program with different subsets of additional type annotations. Those additional type annotations are derived by their type inference. It outputs a program with optimal performance suggested by its analysis, where a subset of additional type annotations are appended to that program. The main difference from our method is an approach to select additional type annotations. Herder performs comprehensive analysis of all possible subsets of additional type annotations with a carefully designed cost analysis, while TypePycker employs a lightweight, amortized approach. Another difference is that Herder integrates type inference with its selection, while our selection is independent of an external type inference engine.

We compare the compilation time of TypePycker and Herder across the benchmark programs. We also compare the execution time of each benchmark program with the selected additional type annotations by TypePycker against that by Herder.

Experiment results illustrate that our method achieves more stable compilation time than Herder. Figure 4 presents the compilation time by Herder and TypePycker across the benchmark programs. Note that some benchmarks are excluded from the comparison because Herder does not support several Python features such as classes. Although Herder shows acceptable and faster compilation time for small programs, it scales poorly to other programs. For example, both Herder and TypePycker give compilation time within 1 second for the programs from MicroBench in Figure 4a. Herder is faster than TypePycker in most of these programs. However, the compilation time by Herder exceeds 1 second in 12 of 28 programs, and exceeds 10 seconds in 7 programs. It takes more than 2000 seconds in the worst case, which is a non-artificial program from SICP. The slow compilation time arises from the large search space that Herder must explore. Although Herder adopts variational typing [9] for efficient computation compared to a brute-force enumeration, the possible number of different occurrences of runtime casts in a program is exponential considering general cases. For example, a program would contain numerous different occurrences of runtime casts under different sets of type annotations for function parameters if the program contains deeply nested function calls. Herder must analyze the costs for all different combinations to suggest optimal performance.

In contrast, our method shows stable compilation time across all benchmark programs. It is less than 1 second in most of the cases including those cases where Herder consumes a significant amount of time. We also measured the elapsed time for type inference and that for inserting selected type annotations separately in the compilation time by TypePycker. Across all the benchmark programs in Figure 4, type inference took 0.23 seconds in average; the later process took 0.026 seconds in average. The higher cost of type inference is because it invokes an SMT solver for constraint solving by the used toolkit InferType as noted in Section 4.1. This relatively heavy cost of invoking an SMT solver is a reason that TypePycker is slower than Herder for MicroBench. In Herder, we cannot measure the type-inference time separately from the selection time since its type inference and the selection process are integrated together.

Experiment results also indicates a tendency that the proposed method can achieve faster and stable compilation time for programs with nested function calls. TypePycker gives significantly faster compilation time than Herder for all the programs from Syn in Figure 4d. The compilation time is less than 1 second for the four programs by TypePycker, while it ranges from 3 to 627 seconds by Herder. At the same time, Chosen is faster than Infer with an up to 6x speedup ratio for these programs in Figure 3d. Although these programs are modified variants based on existing benchmarks, their results provide evidence that there exist programs where our method offers clear advantages. TypePycker shows a win case (Chosen outperforms Infer) while it gives faster compilation time than Herder in 7 of 28 programs

We demonstrate that the proposed method produces comparable execution performance compared to Herder. Figure 5 shows the execution time of the benchmark programs appended with additional type annotations selected by Herder and Type-Pycker. Herder suggests type annotations only for function parameters and function returns. For fair comparison, here we restrict the appended type annotations to only

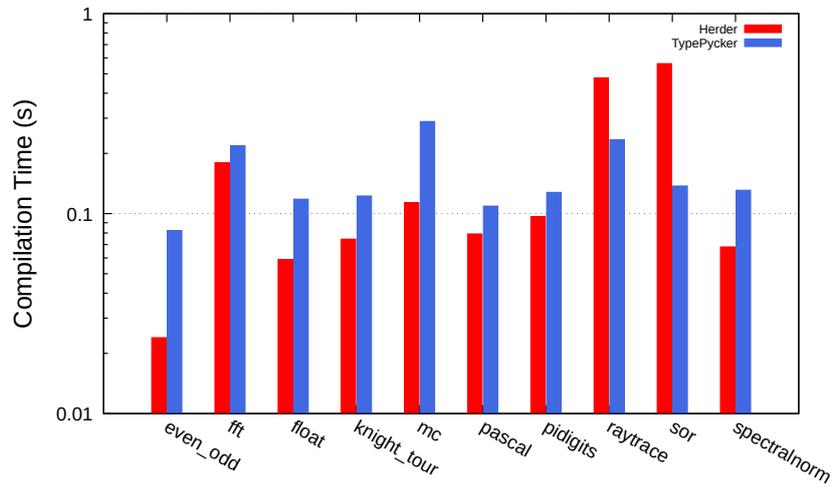

**(a)** MicroBench

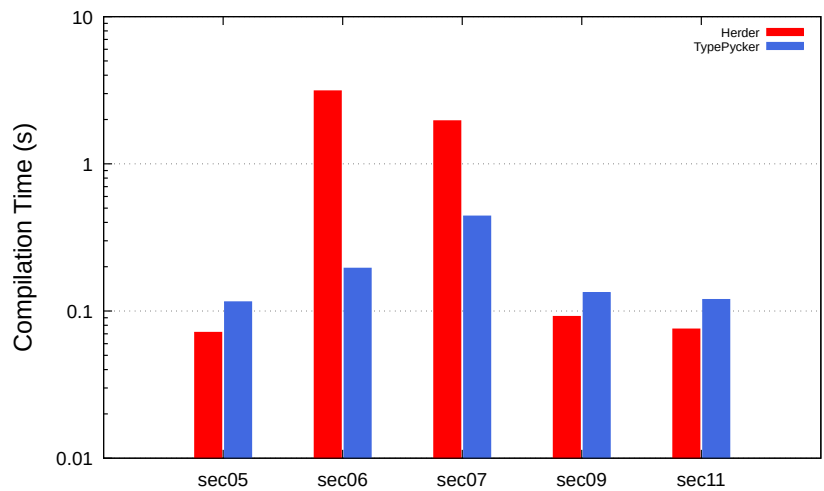

**(b)** PyTextbook

■ **Figure 4**   Compilation time results by Herder and TypePycker (Part I).

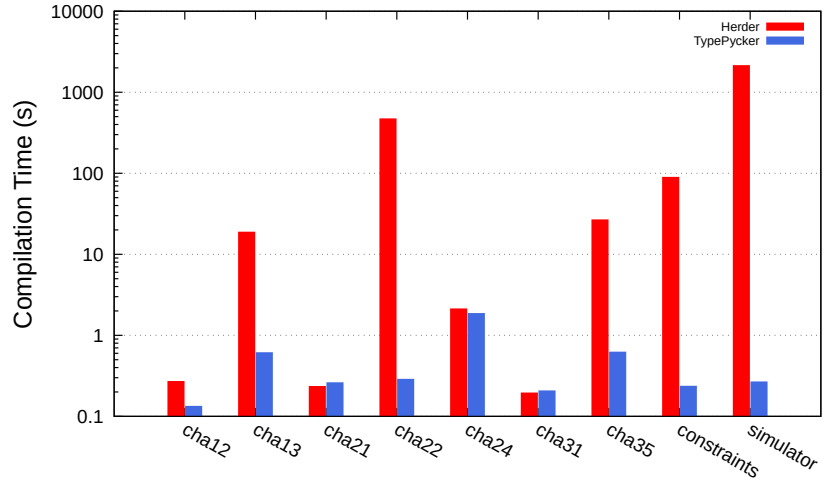

**(c)** SICP

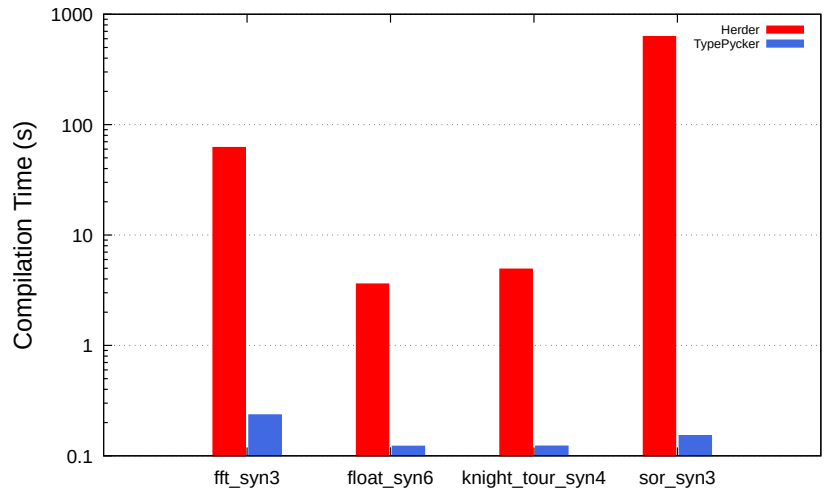

**(d)** Syn

■ **Figure 4** Compilation time results by Herder and TypePycker (Part II). Each sub-figure groups programs by data source. X-axis shows program names. Y-axis shows compilation time in seconds.

**Efficient Selection of Type Annotations for Performance Improvement in Gradual Typing**

function parameters and function returns, excluding ones for local variables. The programs are executed 10 times using our modified Reticulated Python. The only difference in the comparison here is the appended type annotations for function parameters and function returns.

Chosen is faster Herder in 17 of 28 programs, while it is slower than Herder in the remaining 11 programs. The execution time ratio (Herder over Chosen) ranges from 0.5x to 10x. Chosen gives better performance than Herder in several programs mainly because our type inference engine could infer more function parameter types, which enables more optimization opportunities.

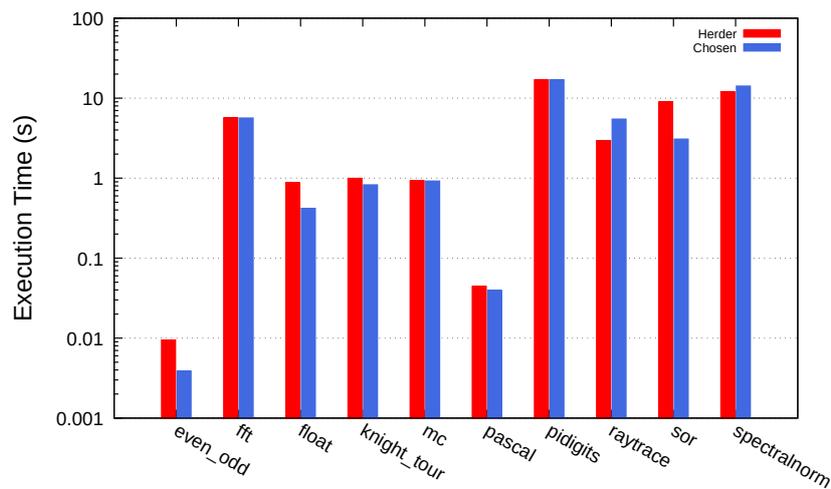

**(a)** MicroBench

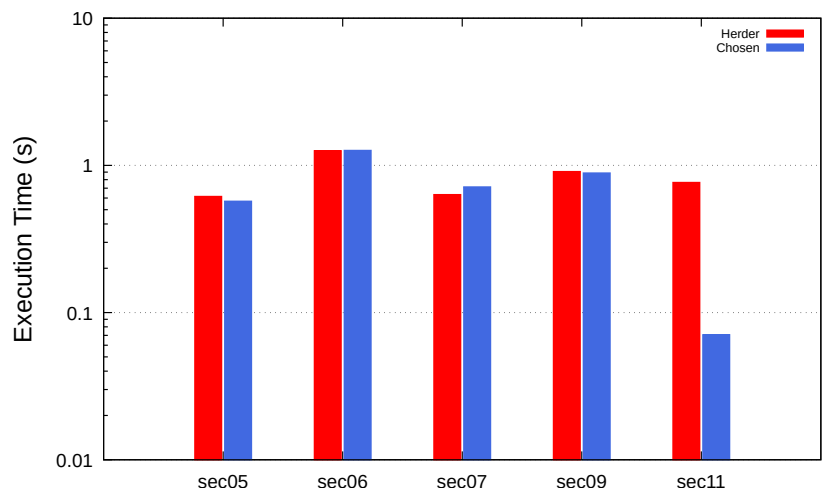

**(b)** PyTextbook

**Figure 5** Execution time results by Herder and TypePycker (Part I).

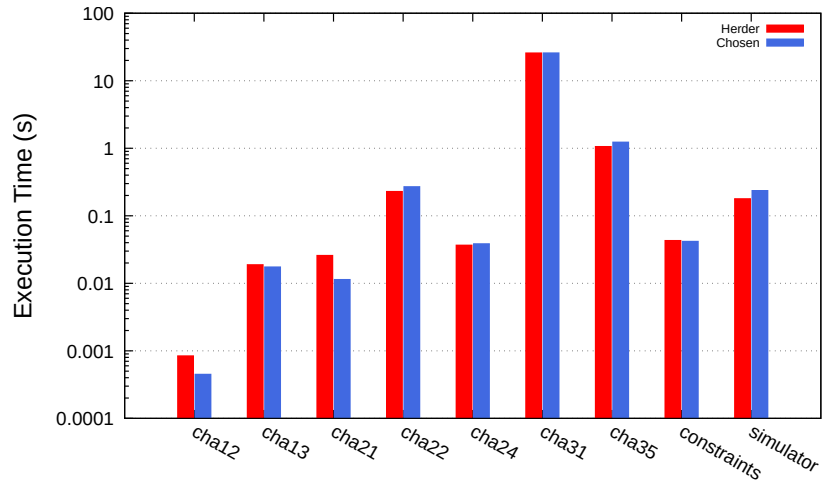

**(c)** SICP

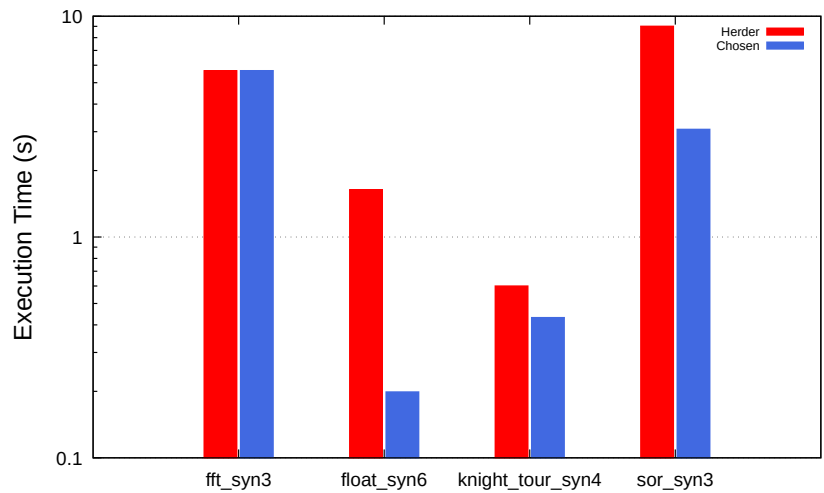

**(d)** Syn

■ **Figure 5** Execution time results by Herder and TypePycker (Part II). Each sub-figure groups programs by data source. X-axis shows program names. Y-axis shows execution time in seconds. Each bar is average of 10 runs.

### 4.4 Assessing Performance Impact across Language Platforms

We demonstrate how the proposed method affects the execution performance by running the benchmark programs on four different language platforms based on Reticulated Python. Here, a language platform refers to a concrete implemented variant of the underlying language Reticulated Python. Each implemented language variant would produce different occurrences of runtime casts given a same program since expressions may be given different static types. We implement the four language platforms by applying different combinations of our two main modifications over the original Reticulated Python: the fast-slow technique and type annotation support for local variables. Note that omitting the fast-slow technique will not preserve program behaviors because additional type annotations will be directly appended to the original functions. The goal of this subsection is to demonstrate that the performance gains are not simply due to our modifications over the original Reticulated Python, but the selection of additional type annotations by our proposal contributes to the observed performance gains.

We compare the execution time of appending only selected type annotations by our method with the execution time of appending all additional type annotations derived from type inference. Each program was executed 10 times on each language platform, and the elapsed time was measured. Same as the comparison in Section 4.2, we categorize the relative execution time of Infer and Chosen into win, tie, loss cases by considering the mean and standard error. The results for 9 benchmark programs are excluded from the comparison below because Chosen appends the same additional type annotations as Infer.

The results indicate that the proposed method can give positive performance impact on four different language platforms. We denote a language platform with support for the fast-slow technique as $F$, support for local variable type annotations as $L$, and no support as $n$. For example, a language platform supporting local variable type annotations but no fast-slow technique is denoted as $nL$. Chosen either outperforms (win), or achieves comparable performance (tie) compared to Infer in 29, 27, 40, and 38 programs across 41 programs on the four language platforms $nn$, $nL$, $Fn$, and $FL$, respectively. The results from the four language platforms indicate that the performance gains are not only from our modifications over Reticulated Python. The selection of type annotations by our method contributes to the performance improvement.

Table 2 demonstrates both categorization results and performance ratios for several selected benchmark programs across language platforms. Those programs are selected to showcase three scenarios: Given a program, our method presents a) either win, tie or loss cases on all language platforms; b) a mix of win/loss and tie cases on all language platforms; c) a win case on one platform but a loss case on another platform.

■ **Table 2** Results of selected programs on four language platforms. Horizontal lines group programs by categorization scenarios. The second column shows the categorization results on four platforms. W, T, L represent win, tie, loss. The later columns show speedup ratios (Infer over Chosen) on each platform. > 1 is better.

| Program | Categorizations | *nn* | *nL* | *Fn* | *FL* |
|---|---|---|---|---|---|
| disjoint_set | WWWW | 1.19 | 1.18 | 1.20 | 1.17 |
| list | WWWW | 2.45 | 2.44 | 1.80 | 1.81 |
| sec07 | WWWW | 1.20 | 1.14 | 1.06 | 1.14 |
| sec09 | WWWW | 1.34 | 1.35 | 2.25 | 2.22 |
| fft | WWTW | 1.19 | 1.15 | 1.00 | 1.16 |
| sieve | TWTW | 0.99 | 1.03 | 0.99 | 1.04 |
| cha13 | TWWT | 1.02 | 1.11 | 1.13 | 0.99 |
| cha24 | LLTT | 0.98 | 0.92 | 0.99 | 1.00 |
| dijkstra_alternate | LLWL | 0.87 | 0.91 | 1.08 | 0.91 |
| knight_tour | LLWW | 0.24 | 0.56 | 1.20 | 3.00 |
| cha31 | WLTL | 1.02 | 0.94 | 1.00 | 0.98 |
| sor_syn3 | LLWW | 0.06 | 0.06 | 2.94 | 2.94 |

### 4.5 Testing Behaviors with JIT

We test several benchmark programs on Reticulated Python with PyPy,[5] a Python variant with JIT compilation, to showcase whether TypePycker can improve the execution performance on a gradually typed language with JIT support. Figure 6 presents the execution time of the tested programs appended with all additional type annotations by type inference and only selected type annotations by TypePycker. We find that TypePycker can achieve performance improvement in some benchmark programs. It is also observed that TypePycker gives worse or different performance impact on PyPy compared to CPython. For example, TypePycker outperforms using all additional type annotations by type inference on CPython in fft and so46582238 (Figure 3). However, TypePycker slows down the execution speed of the two programs on PyPy. Since the comparison is limited and also the number of tested benchmarks is relatively small, a more comprehensive evaluation is needed to draw a reliable conclusion whether TypePycker can benefit a gradually typed language based on JIT or other dynamic-analysis techniques. We regard this as one direction of our future work.

---

[5] Most of the other benchmarks encounter errors on Reticulated Python with PyPy, which is also reported by an existing work [7].

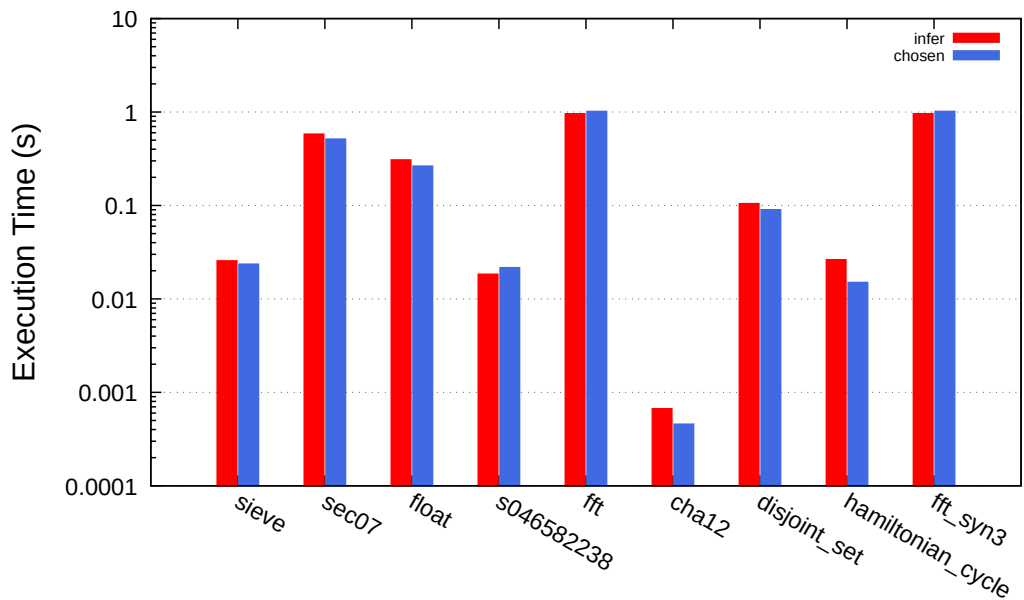

■ **Figure 6** Execution time results on Reticulated Python with PyPy. X-axis shows program names. Y-axis shows execution time in seconds. Each bar is average of 10 runs.

## 5 Threats to Validity

This research work is threatened by several potential factors that could affect the validity of our findings.

This work does not formally discuss the correctness of its optimization with respect to preserving program behaviors. The correctness relies on the underlying type inference engine. To maintain the correctness, one must use a type inference engine only deriving sound types that do not change program behaviors. Exploring such an engine falls outside the scope of this paper. We mitigate this threat in two limited ways. First, we manually validate the correctness of the inferred types for benchmark programs in the experiments. We confirm that they do not change the behaviors of the benchmark programs. Second, we test our implementation with several challenging programs written in a recent research paper [7]. Those programs are specifically designed to test the correctness. Our engine successfully preserves the behaviors of those programs. We include those programs in our artifact.

The performance gains observed in the experiments might be more pronounced than what would be achieved in practical scenarios. This limitation arises primarily from using the collected un-annotated programs as our baseline. Although those benchmark programs are treated partially typed when executed by Reticulated Python in the experiments, real-world programs may be already annotated with some degree of concrete type annotations for function parameters, function returns or local variables by programmers in a gradually typed language. We were not able to collect a suitable dataset with real annotated programs.

## 6 Related Work

Understanding the performance changes in gradual typing has become a significant research area following the development of a wide range of gradually typed languages. Existing gradually typed languages are mainly classified into two broad categories: erasure and non-erasure semantics [10, 13]. Languages with erasure semantics allow values of unexpected types to flow from untyped code into typed code by erasing types at runtime. Examples are TypeScript, Flow and mypy. Languages with non-erasure semantics, in contrast, enforce static types at runtime by performing runtime casts at the boundary between untyped and typed code. They are further distinguished between deep semantics such as Gradualtalk [2], Typed Racket [42] and first-order semantics such as Reticulated Python [43], Nom [34]. Several studies have collectively discovered significant runtime overhead in gradually typed languages with non-erasure semantics in such as Typed Racket [14, 41] and Reticulated Python [6, 15]. In this paper, we do not consider languages such as TypeScript since programs written under erasure semantics would not impose significant runtime overhead from static type annotations. The goal of this work is to develop a compile-time technique for mitigating the performance penalty found in gradually typed languages with non-erasure semantics. The investigated language Reticulated Python in our experiments is one instance. Exploring the effectiveness of the proposal on other gradually typed languages [22, 31, 42] with non-erasure semantics is our future work.

A number of research efforts have emerged following the report of performance degradation in gradual typing. One major direction is to enhance static information for improving the performance. Rastogi et al. [37] firstly used type inference to optimize gradually typed programs on ActionScript [33] by considering a correctness goal of preserving program behaviors. Vitousek et al. [44] optimized the transient semantics of Reticulated Python by removing runtime casts verified by type inference. Following the principle of preserving program behaviors, Campora et al. [7] proposed discriminative typing to safely improve the performance by keeping both an optimized version of a function and also the unoptimized original function based on type inference. Our implementation adopts Campora et al.'s idea of maintaining both optimized and unoptimized functions. Unlike their approach, however, our method does not formally guarantee that a runtime error is blamed at the same location because we use a different type system. Our method differs in selectively adding type annotations derived by type inference for better performance.

There are research studies for addressing the performance pitfalls in gradual typing by statically reasoning how types affect the performance. Herder [6] is a static cost analysis tool to estimate the execution time of different combinations of typed parameters within a given program. These combinations are often termed *configurations* in the context of gradual typing. The tool outputs a configuration expected to yield optimal execution time based on its cost analysis. Our novelty compared to Herder is an efficient method for selecting additional type annotations. TypePycker does not need to analyze all possible configurations. Experiment results showed that TypePycker can achieve comparable execution performance while having a more stable compilation time. Khan et al. [21] developed a machine-learning based approach *LearnPerf* to predict the performance of possible configurations in a given program. Similar to TypePycker, LearnPerf uses an external type inference engine to derive additional type annotations. Besides the difference presented in Section 2, LearnPerf differs from TypePycker in two ways. First, LearnPerf does not consider preserving program behaviors. Second, LearnPerf trains an individual model for each input program, which limits its portability for practical usage.

Another direction for understanding and optimizing the performance of gradual typing is the use of runtime techniques. Bauman et al. [3] showed practical performance improvement by using a tracing JIT after the report of significant runtime overhead by Takikawa et al. Richards et al. [38] bridged the gap between a virtual machine leveraging JIT compilation and gradually typed languages to eliminate redundant runtime casts. Roberts et al. [39] employed a standard JIT compilation to reduce the overhead of transient runtime casts. This paper primarily examines the proposed method on Reticulated Python without additional runtime techniques such as JIT since we aim to study the genuine effect of the proposal as a compile-time technique. Although the experiments testing with PyPy in Section 4.5 are rather limited, we believe that the proposed method can be integrated and potentially give performance improvement to other offline compilation and runtime techniques. Greenman et al. [14] and Hejduk et al. [17] explored how profiling data can be leverated to navigate decisions for type migration by a large-scale empirical study based on the rational-programmer [23, 24] method. Our work targets a different point in providing a compile-time technique to automatically improve the performance of a gradually typed program.

Static type inference has been a well-established technique across a variety of programming paradigms. Type inference benefits fundamental applications such as early error detection [19, 29] and performance improvement [8]. In Python, there are numerous existing type inference engines leveraging diverse techniques such as rule-based systems [12], abstract interpretation [32], SMT solvers [16, 45], and machine-learning based approaches [18, 30, 35]. In this work, we implement a custom type inference engine using a compiler toolkit InferType [26]. Although we have not empirically explored the integration with existing engines, we expect that the proposed method can be fluently extended since it is independent of the process of type inference.

Several data flow analyses have been applied in the area of type inference. Maggi et al. [27] presented an approach to infer types related to Java bytecode using a data flow analysis. Zhou et al. [46] proposed a method of type inference based on a data flow analysis for decompilation. Bhanuka et al. [4] captured the data flow by using subtype constraints for generating readable type error messages in constraint-based type inference systems. Our proposal shares similarities with those data flow techniques in the construction of a graph for modeling value propagation. Our contribution is distinct in selectively adding type annotations for improving the performance of gradually typed programs.

## 7  Conclusion

We propose TypePycker, a new lightweight method to efficiently select type annotations derived by type inference for improving the execution performance of gradually typed programs. As significant costs of runtime casts often occur when a value repeatedly crosses the boundaries between untyped to typed code, the proposal selects type annotations placed along the data flows. It quickly checks whether all untyped variables along the data flows are expected to obtain type annotations derived by type inference using a lightweight, amortized approach.

We demonstrated the usage of TypePycker, and experimented its effectiveness in improving the execution performance on Reticulated Python. We found that Type-Pycker could achieve better execution speed than simply using all the inferred types derived by type inference in the benchmark programs with a speedup ratio up to more than 5x. In comparison with an existing approach, TypePycker delivered comparable execution performance with more stable compilation time. We also observed that TypePycker achieved much faster compilation time in real-world programs where the existing approach incurred significantly long compilation time such as more than 10 minutes. We believe that the proposal is a candidate approach for helping mitigate the performance bottleneck of gradual typing in practical scenarios.

### Data-Availability Statement

The artifact associated with this paper is available on Zenodo [25].

**Acknowledgements**  The work is partly supported by JSPS KAKENHI JP24H00688.

**About the authors**

**Senxi Li** Contact at lisenxi@csg.ci.i.u-tokyo.ac.jp.
https://orcid.org/0009-0008-2644-7763

**Feng Dai** Contact at daifeng@csg.ci.i.u-tokyo.ac.jp.
https://orcid.org/0009-0006-0995-2536

**Tetsuro Yamazaki** Contact at yamazaki@csg.ci.i.u-tokyo.ac.jp.
https://orcid.org/0000-0002-2065-5608

**Shigeru Chiba** Contact at chiba@csg.ci.i.u-tokyo.ac.jp.
https://orcid.org/0000-0002-1058-5941